\documentclass[prb,a4paper,twocolumn,floatfix,showpacs,showkeys,amsmath,amssymb,nobibnotes,altaffilletter]{revtex4}

\usepackage{graphicx}% Include figure files
\usepackage{pslatex}

\begin{document}

\title{Polaron Recombination in Pristine and Annealed Bulk Heterojunction Solar Cells}

\author{C.~Deibel}\email{deibel@physik.uni-wuerzburg.de}
\affiliation{Experimental Physics VI, Julius-Maximilians-University of W{\"u}rzburg, D-97074 W{\"u}rzburg}

\author{A.~Baumann}
\affiliation{Experimental Physics VI, Julius-Maximilians-University of W{\"u}rzburg, D-97074 W{\"u}rzburg}

\author{V.~Dyakonov}
\affiliation{Experimental Physics VI, Julius-Maximilians-University of W{\"u}rzburg, D-97074 W{\"u}rzburg}
\affiliation{Functional Materials for Energy Technology, Bavarian Centre for Applied Energy Research (ZAE Bayern), D-97074 W{\"u}rzburg}

\date{\today}

\begin{abstract}

We determined the dominant polaron recombination loss mechanism in pristine and annealed polythiophene:fullerene blend solar cells by applying the photo-induced charge extraction by linearly increasing voltage (photo-CELIV) method in dependence on temperature. In pristine samples, we find a strongly temperature dependent bimolecular polaron recombination rate, which is reduced as compared to the Langevin theory. For the annealed sample, we observe a polaron decay rate which follows a third order of carrier concentration almost temperature independently.

\end{abstract}



\maketitle

Solution processed organic solar cells have reached efficiencies of 5--6\% power conversion efficiency recently.\cite{green2008,laird2008} In order to further optimize these devices, a better understanding of the fundamental loss mechanisms is required, including charge carrier recombination. A versatile experimental technique to study charge transport properties in organic electronic devices is photo-CELIV. It enables the simultaneous investigation of charge carrier density, mobility and lifetime by recording transient currents in real devices, which sets it apart from other experiments. In this study, we report temperature-dependent photo-CELIV measurements on poly(3-hexyl thiophene):[6,6]-phenyl-C61 butyric acid methyl ester (P3HT:PCBM) bulk heterojunction solar cells. We find reduced polaron decay rates as compared to Langevin recombination, with decay dynamics to the second order for pristine samples, and to the third order for annealed devices. We show that the present theories to explain the reduced recombination rate are not able to describe the experimental temperature dependence.

The classic theory to describe polaron losses in disordered low mobility materials is the Langevin recombination,\cite{pope1999} in which the actual bimolecular recombination process is limited by two charges finding one another by diffusion. The Langevin recombination rate $R_\text{Langevin}$ is therefore proportional to the diffusivity and thus the charge carrier mobility $\mu$, 
\begin{equation}
	R_\text{Langevin} = \gamma np
	\label{eqn:L}
\end{equation}
with the Langevin prefactor
\begin{equation}
	\gamma = \frac{q}{\epsilon} \mu .
	\label{eqn:gamma}
\end{equation}
$n$ and $p$ are the negative and positive polaron densities, respectively, $q$ is the elementary charge, and $\epsilon$ the material dielectric constant. Several authors have found the charge recombination rate in disordered materials to be reduced as compared to the Langevin theory, for amorphous Silicon~\cite{tyutnev1992} as well as for polymers.\cite{pivrikas2005a,juska2006} Different theories have been presented in order to explain these experimental findings. Adriaenssens and Arkhipov\cite{adriaenssens1997}  developed a model describing an energy activated recombination process due to the spatial separation of the two charges, governed by potential fluctuations. The effective recombination rate $R_\text{eff}$ is given by
\begin{equation}
	R_\text{eff} = \frac{2\pi\Delta}{kT}\exp\left( -\frac{2\Delta}{kT}  \right) R_\text{Langevin}
	\label{eqn:Reff-aa}
\end{equation}
with $\Delta$ being the activation energy for the the charges to overcome the spatial separation, and $kT$ is the thermal energy. Another model by Koster et al.\cite{koster2006} is also based on the notion of spatial separation, this time in phase separated systems with two components, with the positive and negative charges being restricted to their respective phases. Thus, the authors replace the mobility in the Langevin prefactor, typically, the spatial average of electron and hole mobility, by the minimum mobility of the two charges. We will compare our experimental findings with these two models.

We prepared bulk heterojunction solar cells by spin coating P3HT:PCBM blends made from a solution of (20mg P3HT+20mg PCBM) per ml Chlorobenzene, on poly(3,4-ethylenedioxythiophene) poly(styrenesulfonate) covered ITO/glass substrates. The active layer was about 105nm thick. Aluminum anodes were evaporated thermally. The annealed samples were treated for 10 minutes at 140$^{\circ}$C. P3HT was bought from Rieke Metals, PCBM from Solenne. All materials were used without further purification. All preparation steps were performed in a nitrogen glovebox and an attached thermal evaporation chamber (vacuum).

We investigated the pristine and annealed samples by the photo-CELIV method. Singlet excitons were generated by a short Nitrogen laser pulse ($\lambda$=500nm by dye unit, 5ns, 500$\mu$J/cm$^2$). We assume a very efficient polaron pair generation in the 1:1 blend. After a delay time at zero internal field during which recombination can occur, the remaining polarons were extracted by a triangular voltage pulse in reverse bias. From the extraction current, charge carrier mobility and concentration of extracted charge carriers were obtained simultaneously at different delay times, in analogy to Ref.~\cite{juska2006}. The temperature was varied using a Helium closed-cycle cryostate with contact gas.

\begin{figure}
	\includegraphics[width=7cm]{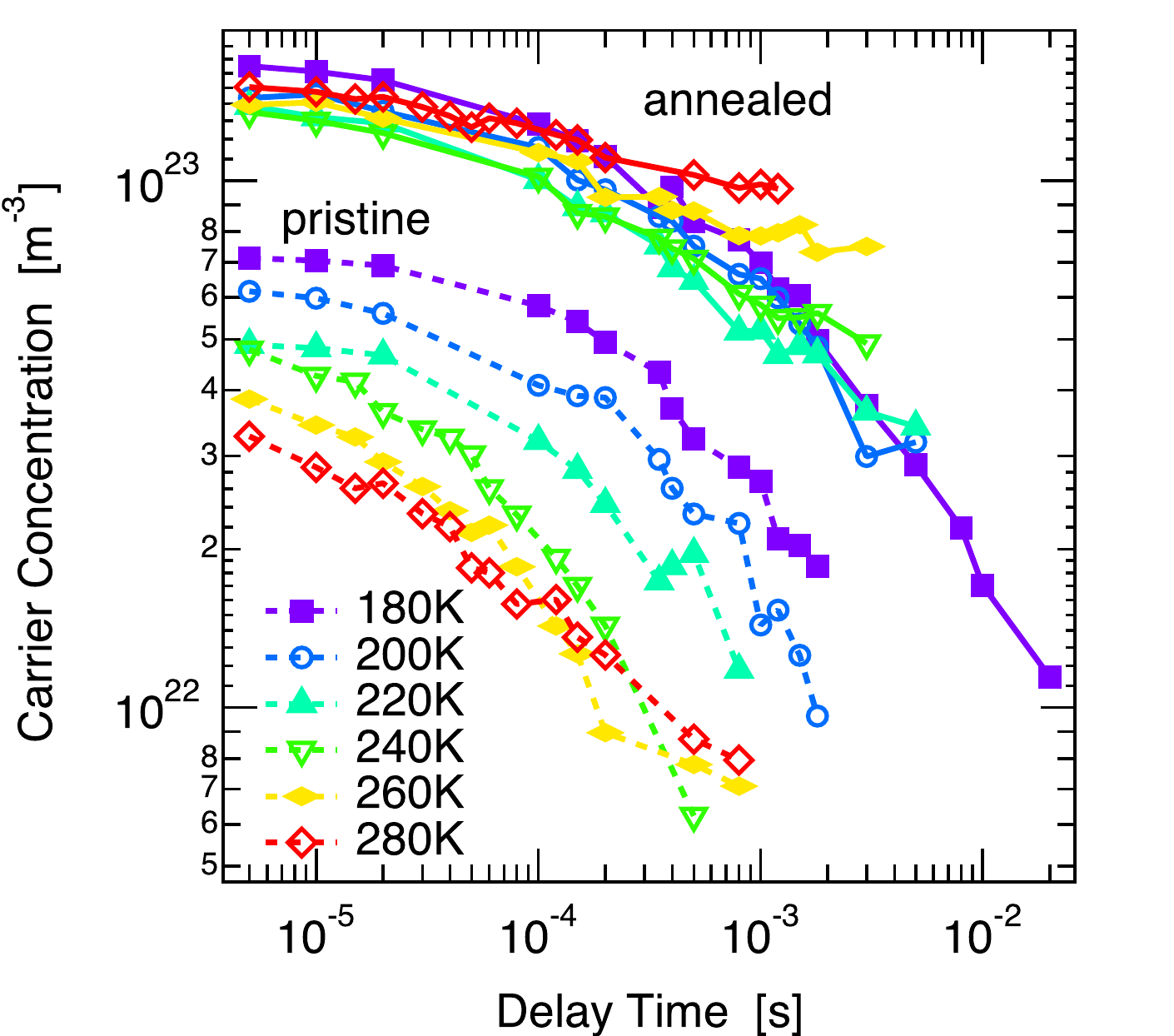}
	\caption{(Color online) Carrier concentration vs.\ delay time of a pristine (bottom) and annealed (top) P3HT:PCBM solar cell for different temperatures.  The pristine sample shows a strong temperature dependence typical for bimolecular Langevin-type recombination. In contrast, the annealed sample shows only a very weak temperature dependence. At long delay times and high temperatures, artefacts due to injection can be observed by the inflexion point.\label{fig:conc_T}}
\end{figure}

In Fig.~\ref{fig:conc_T}, the polaron concentration in dependence on the delay time is shown for various temperatures. The pristine sample generally shows lower concentrations as compared to the annealed sample. Also, the temperature dependence in the pristine sample indicates Langevin-type recombination: at low temperatures, the recombination rate is low due to the low charge carrier mobilities. In contrast, the annealed sample shows almost no temperature dependence. 

\begin{figure}
	\includegraphics[width=7.5cm]{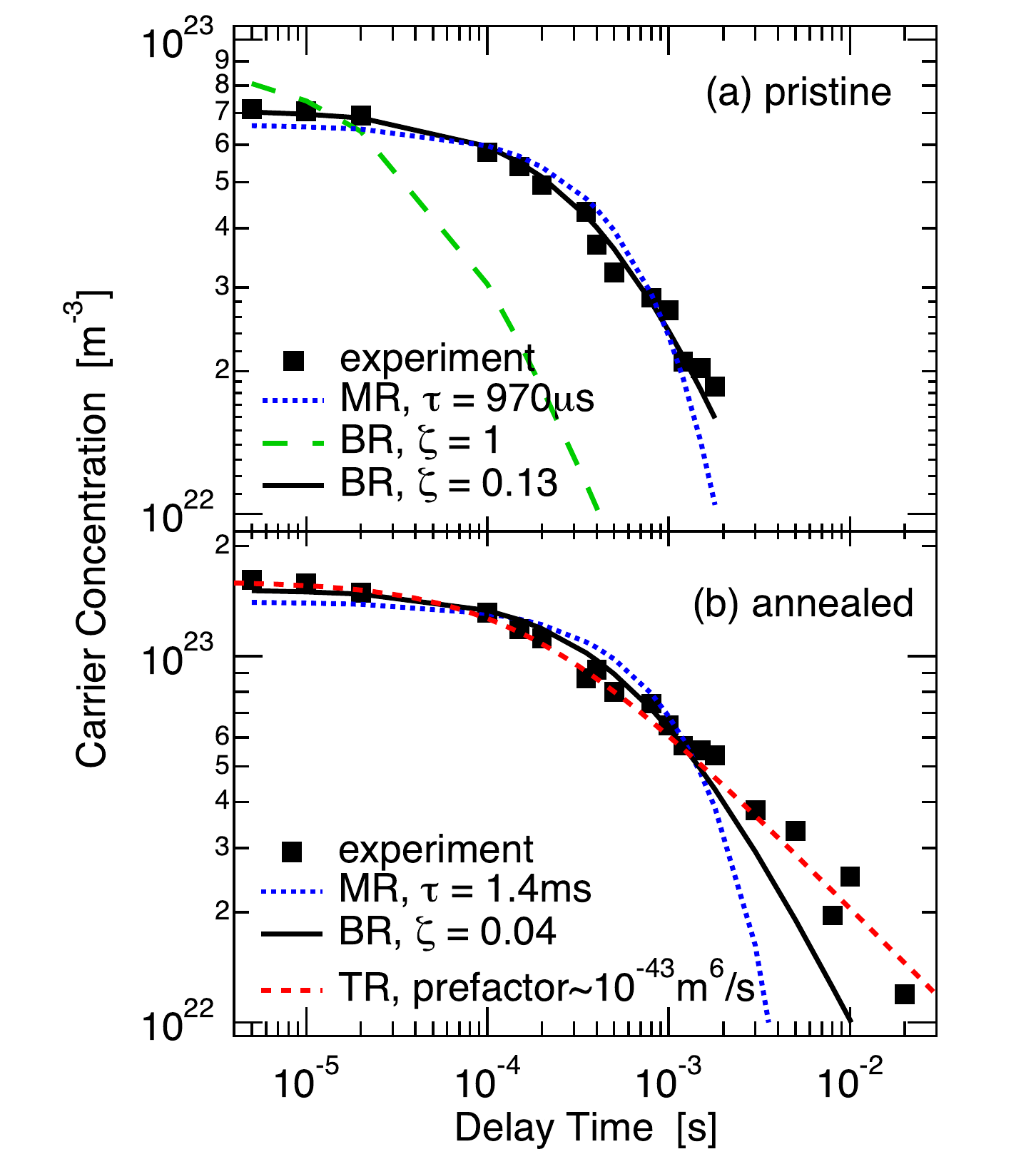}
	\caption{(Color online) Carrier concentration vs.\ delay time of a (a) pristine and (b) annealed P3HT:PCBM solar cell at 180K, extracted from a photo-CELIV experiment.  Numeric fits of the data using the polaron continuity equation are included, namely the recombination types monomolecular (MR, blue dots), bimolecular with Langevin (BR, green dashed line) and reduced Langevin (black solid line) as well as third order (TR, red dashed line).\label{fig:conc_180K}}
\end{figure}

Analysing the spectra at 180K temperature (Fig.~\ref{fig:conc_180K}) in more detail, the decay dynamics can be obtained by fitting to the numerically solved continuity equation for polarons. For monomolecular recombination, the lifetime $\tau$ is determined by the fit. In contrast, the classic Langevin recombination has no fit factor, as we used the time-resolved mobility, which was experimentally determined by photo-CELIV at the same time as the carrier concentration. For the pristine samples, Fig.~\ref{fig:conc_180K}(a), neither monomolecular nor Langevin recombination, nor a combination of both can describe the experiment properly. Only a reduced Langevin recombination $R_\text{eff}=\zeta R_\text{Langevin}$, can capture the experiment, with a dimensionless reduction factor $\zeta$ smaller than one. For the annealed sample, where we can measure to longer decay times due to a better signal-to-noise ratio, a reduced Langevin recombination gives a reasonable fit. However, the best fit is obtained using a third order decay law.
Such a decay has recently been reported by Shuttle et al.\cite{shuttle2008} for annealed polythiophene:fullerene solar cells. They used  transient photovoltage and transient absorption spectroscopy, concluding a third order recombination from light intensity dependent measurements. Our measurements, showing the third order decay directly, are thus a verification of these findings.

\begin{figure}
	\includegraphics[width=7cm]{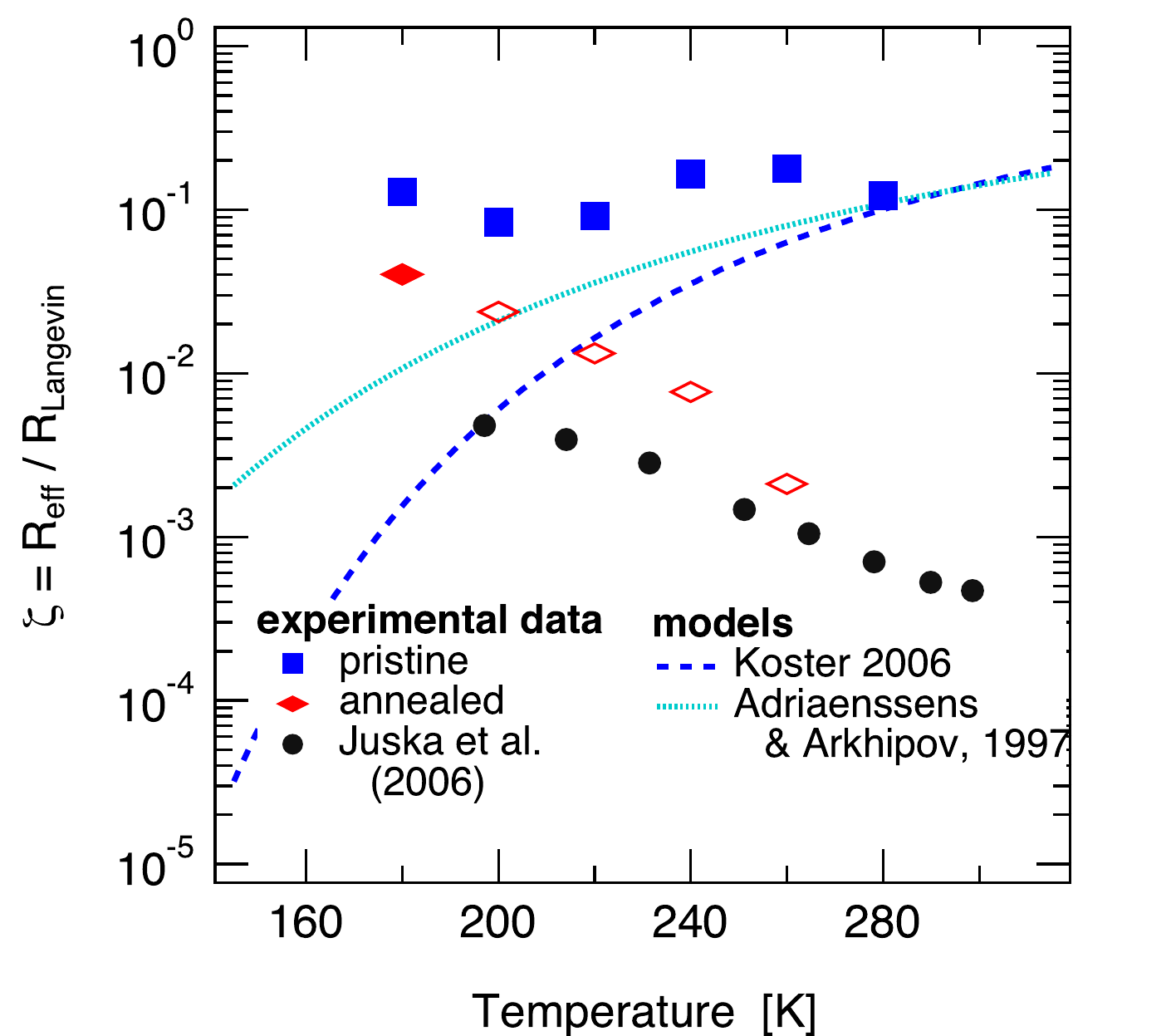}
	\caption{(Color online) Temperature dependence of the Langevin reduction factor $\zeta$, the fraction of effective over Langevin recombination. Shown are the results for pristine (squares) and annealed (diamonds; the open symbols signify the determination of $\zeta$ ignoring the parasitic injection). They are compared to values from Juska et al.\cite{juska2006} (circles). The current models for the description of the reduced Langevin recombination from Koster et al.\cite{koster2006} as well as Adriaenssens  and Arkhipov\cite{adriaenssens1997} --- showing a temperature dependence not observed in the experiments --- are also included.\label{fig:zeta_T}}
\end{figure}

We have assembled the temperature-dependent Langevin reduction factors $\zeta=R_\text{eff}/R_\text{Langevin}$ in Fig.~\ref{fig:zeta_T}. For fitting the bimolecular recombination For the pristine samples, $\zeta$ is constant within a narrow range (related to the strong temperature dependence of the decay signal). For comparison, we included also $\zeta$ of the annealed samples, in spite of (i) the weak temperature dependence of the charge carrier transients shown in Fig.~\ref{fig:conc_180K}, and (ii) the fit to the experimental data being better with a third order decay law. Thus, we point out that the apparent strong temperature dependence of $\zeta$ for the annealed sample is due to the worse fits at higher temperatures, as the charge carrier transients in Fig.~\ref{fig:conc_180K} then become subsequently shorter. As the transients of the annealed sample generally show almost no temperature dependence, the value at 180K is of the highest relevance due to its best signal-to-noise ratio and lowest dark injection. Find for comparison also the data from Juska et al.\cite{juska2006} for the same material system.

The models by Adriaenssens et al.,\cite{adriaenssens1997} Eqn.~(\ref{eqn:Reff-aa}), as well as Koster et al.\cite{koster2006} are included in Fig.~\ref{fig:zeta_T} with typical parameters. Both show temperature dependent Langevin reduction factors $\zeta$ markedly different from our experimental results. The spatial separation of positive and negative polarons in the polymer:fullerene blends will certainly have an influence on the charge carrier recombination, nevertheless does neither model capture the experimental temperature dependence of $\zeta$. 

A comparison of the theoretical models and our experimental findings allows to draw several conclusions. The pristine polymer:fullerene solar cell can be described with a reduced Langevin recombination, according to the experimental carrier dynamics and their temperature dependence. The resulting Langevin reduction factor is almost temperature independent, and thus not in accordance to the models of Adriaenssens et al.\cite{adriaenssens1997} and Koster et al.\cite{koster2006} We note that we cannot measure beyond a delay time of about 2ms for the pristine sample due to a low signal-to-noise ratio. As a fit for third order decay shows a significant difference to second order decay only at longer time scales, we cannot completely rule out third order polaron decay for the pristine samples yet, even though the temperature dependence is rather typical for Langevin-type recombination. In contrast, the annealed sample shows a charge carrier recombination which is almost temperature independent, opposite to the expectations for Langevin-type recombination. Also, the direct decay can best be fitted with a third order process. As Shuttle et al.\cite{shuttle2008} observed similar findings using complementary experimental methods, and investigating the illumination density rather than temperature dependence, we see our results as verification. The question arising now is the origin of this third order decay law. In principle, either a bimolecular recombination with carrier concentration (or time dependent) prefactor is possible, as is a trimolecular recombination process. Ref.~\cite{shuttle2008} tends to the first explanation, which seems indeed more probable. The carrier concentration dependence of the recombination prefactor could stem from either a time dependent mobility due to carrier thermalisation, a carrier concentration dependence of the charge carrier mobility, or a carrier concentration dependence of the (as of yet) physically unspecific prefactor $\zeta$. The first of these options is unlikely, as the photo-CELIV technique allows direct insight into carrier thermalisation; it is therefore already included in our analysis. The latter two options raise the question of the weak temperature dependence of the carrier recombination. Considering trimolecular recombination, general candidates might be either Auger-like processes or a recombination of trions~\cite{kadashchuk2007}, a proposed metastable system of two polarons of the same charge on the polymer chain bound to a trapped polaron on the fullerene. Nevertheless, as both trimolecular processes are unexpected, this topic needs further investigations due to the implications on the fundamental understanding of organic semiconductors and the modelling of polymer:fullerene solar cells. 

In conclusion, applying time-resolved photo-CELIV experiments on polythiophene:fullerene solar cells, we investigate the polaron recombination dynamic and its importance as charge carrier loss mechanism. We observe a bimolecular polaron recombination, the decay rate and temperature dependence indicating a reduced Langevin-type process for pristine samples. 
For annealed solar cells, the polaron decay shows a third order polaron decay. The temperature dependence of the recombination mechanism is distinctly different from the pristine sample, being almost temperature independent. Monomolecular recombination processes were not observed. 

\begin{acknowledgments}

A.B. thanks the Deutsche Bundesstiftung Umwelt for funding. V.D.'s work at the ZAE Bayern is financed by the Bavarian Ministry of Economic Affairs, Infrastructure, Transport and Technology. 

\end{acknowledgments}

%\clearpage \newpage


\begin{thebibliography}{10}

\bibitem{green2008}
{M.~A.} {Green},
  {K.}~{Emery},
  {Y.}~{Hishikawa},
  {and} {W.}~{Warta},
{Prog. Photovolt.} \textbf{{16}},
  {435} ({2008}).

\bibitem{laird2008}
{D.}~{Laird},
  {S.~S.} {Vaidya},
  {S.}~{Jia},
  {S.~B.} {Li}, {and}
  {J.}~{Bernkopf},
  {presented at the SPIE Optics{\&}Photonics 2008, San Diego,
  USA}.

\bibitem{pope1999}
{M.}~{Pope} {and}
  {C.~E.} {Swenberg},
  \emph{Electronic Processes in Organic Crystals and Polymers}
  ({Oxford University Press}, {USA},
  {1999}), {2nd} ed.

\bibitem{tyutnev1992}
{A.~P.} {Tyutnev},
  {A.~I.} {Karpechin},
  {S.~G.} {Boev},
  {V.~S.} {Saenko},
  {and} {E.~D.}
  {Pozhidaev}, {Phys. Stat. Sol. A}
  \textbf{{132}}, {163} ({1992}).

\bibitem{pivrikas2005a}
{A.}~{Pivrikas},
  {G.}~{Ju{\v{s}}ka},
  {A.~J.} {Mozer},
  {M.}~{Scharber},
  {K.}~{Arlauskas},
  {N.~S.} {Sariciftci},
  {H.}~{Stubb}, {and}
  {R.}~{{\"O}sterbacka},
{Phys. Rev. Lett.} \textbf{{94}},
  {176806} ({2005}).

\bibitem{juska2006}
{G.}~{Ju{\v{s}}ka},
  {K.}~{Arlauskas},
  {J.}~{Stuchlik}, {and}
  {R.}~{{\"O}sterbacka},
{J. Non-Cryst. Sol.} \textbf{{352}},
  {1167} ({2006}).

\bibitem{adriaenssens1997}
{G.~J.} {Adriaenssens}
  {and} {V.~I.}
  {Arkhipov}, {Sol. State Comm.}
  \textbf{{103}}, {541} ({1997}).

\bibitem{koster2006}
{L.~J.~A.} {Koster},
  {V.~D.} {Mihaletchi},
  {and} {P.~W.~M.}
  {Blom}, {Appl. Phys. Lett.}
  \textbf{{88}}, {052104}
  ({2006}).

\bibitem{shuttle2008}
{C.}~{Shuttle},
  {B.}~{O'Regan},
  {A.}~{Ballantyne},
  {J.}~{Nelson},
  {D.}~{Bradley},
  {J.~D.} {Mello}, {and}
  {J.}~{Durrant},
{Appl. Phys. Lett.} \textbf{{92}},
  {093311} ({2008}).

\bibitem{kadashchuk2007}
{A.}~{Kadashchuk},
  {V.~I.} {Arkhipov},
  {C.-H.} {Kim},
  {J.}~{Shinar},
  {D.-W.} {Lee},
  {Y.-R.} {Hong},
  {J.-I.} {Jin},
  {P.}~{Heremans}, {and}
  {H.}~{B{\"a}ssler},
{Phys. Rev. B} \textbf{{76}},
  {235205} ({2007}).

\end{thebibliography}
\end{document}